# *In-situ* Raman spectroscopy analysis of the interfaces between Ni-based SOFC anodes and stabilized zirconia electrolyte

D.A. Agarkov*, I.N. Burmistrov, F.M. Tsybrov, I.I. Tartakovskii, V.V. Kharton and S.I. Bredikhin

Institute of Solid State Physics RAS, 2 Academician Osipyan Str., Chernogolovka 142432, Moscow Distr., Russia

* Corresponding author. Tel: +7 496 522 83 38; E-mail: agarkov@issp.ac.ru

**Abstract**

A new experimental approach for *in-situ* Raman spectroscopy of the electrode | solid electrolyte interfaces in controlled atmospheres, based on the use of optically transparent single-crystal membranes of stabilized cubic zirconia, has been proposed and validated. This technique makes it possible to directly access the electrochemical reaction zone in SOFCs by passing the laser beam through single-crystal electrolyte onto the interface, in combination with simultaneous electrochemical measurements. The case study centered on the analysis of NiO reduction in standard cermet anodes under open-circuit conditions, demonstrated an excellent agreement between the observed kinetic parameters and literature data on nickel oxide. The porous cermet reduction kinetics at 400-600°C in flowing $H_2$-$N_2$ gas mixture can be described by the classical Avrami model, suggesting that the reaction rate is determined by the metal nuclei growth limited by Ni diffusion. The advantages and limitations of the new experimental approach were briefly discussed.

## 1. Introduction

Redox kinetics and local variations in the charge carrier concentration gradients across the electrode | electrolyte interface are among performance-determining factors of solid oxide fuel cell (SOFC) electrodes [1-3]. Although microscopic mechanisms of the electrode reactions can be evaluated by well-known electrochemical methods, the resultant information is usually cumulative, except for the microelectrode techniques. One important complementary technique enabling real-time analysis of the SOFC reactions is the Raman spectroscopy, successfully applied for a variety of model systems, processes and fuel cell materials [4-13]. Due to low penetration depth of the excitation radiation, however, most approaches known in the literature (e.g., [5-13]) are mainly based on the Raman spectra collected from the outer boundaries of model electrochemical cells, primarily surfaces of the electrodes and solid electrolyte membranes. The electrochemical reaction zones where the majority of ionic and electronic charge carriers are generated, such as the area in vicinity of the triple-phase boundary (TPB), can hardly be achieved to a sufficient extent by viewing the surfaces and edges of the electrode systems.

Following our previous reports on the developments and validation of a new combined technique for *in-situ* Raman spectroscopy of the electrode | solid electrolyte interfacial zones under SOFC operating conditions [14,15], the present work is centered on the studies of reduction kinetics of standard Ni-based cermet anodes applied onto optically transparent single-crystal membranes of 10 mol.% $Sc_2O_3$ and 1% mol.% $Y_2O_3$ stabilized zirconia (10Sc1YSZ). An appropriate selection of the electrodes geometry and solid electrolyte (SE) membranes makes it possible to directly collect Raman spectra from the TPB zone, passing the beam through single-crystal electrolyte onto the anode | electrolyte interface. The results, briefly summarized in the present work, correspond to the open-circuit conditions; their comparison with the behavior of polarized anode layers and separate anode material will be reported elsewhere.

## 2. Experimental

In order to explain operation principles of the experimental setup elaborated for *in-situ* Raman spectroscopy analysis of the interfaces as a function of temperature, atmosphere and current density, Fig.1 compares the planar SOFC electrode configuration (a) with that of the model electrochemical cells used in this work (b). The working electrode (WE) made of Ni-based cermet for these case studies, has a standard circular shape; the ring-shape counter electrode (CE) enables penetration of the laser beam through the transparent SE membrane. The planar electrochemical cells were hermetically sealed onto a single-crystal sapphire tube, with two thermocouples and Pt current collectors (Fig.1c). Fig. 1d displays general scheme of the entire setup, which comprises a gas-mixing system equipped with mass-flow controllers (MFC, Bronkhorst), high-temperature chamber (Fig.1c), and an optical system. Raman scattering in the WE was excited by a 30 mW green (532 nm) laser. The Raman spectra presented in this work were obtained as a sum of 100 spectra collected for 0.5-2 s each, depending on temperature; the background spectra were automatically subtracted. The equipment and measurement procedures were described in previous publications [14,15].

The transparent SE membranes (thickness of 250 and 500 μm) were cut from the 10Sc1YSZ single crystals, grown by the direct melt crystallization technique in a cold crucible at the Institute of General Physics RAS [16,17]. The fluorite-type cubic structure of the 10Sc1YSZ crystals was confirmed by the Laue method; one pattern is presented in Fig.2a. Optical transparence of the polished membranes was assessed measuring their transmittance spectra (Specord M40 spectrophotometer, Carl Zeiss Jena). In the wavelength region of 540-600 nm, the transmittance was about 70%, high enough to obtain Raman spectra from the WE | SE interface.

The Ni-based cermet anode composition, thickness, morphology and deposition route were optimized in previous works [18,19]. Submicron NiO powder (Sigma Aldrich) was pre-annealed in air at 700°C for 2 h to remove overstoichiometric oxygen and absorbed water [20]. A mixture of pre-annealed NiO (40 wt.%) and 10 mol.% $Sc_2O_3$ + 1 mol.% $CeO_2$ co-stabilized zirconia (10Sc1CeSZ, DKKK, Japan, 60 wt.%) was ball-milled and mixed with organic additives to prepare pastes for screen-printing. The porous anodes were deposited onto one side of the SE disks, dried at 130°C, and then sintered in air at 1250°C for 3 h. Reduction of the anodes was performed in the course of Raman measurements. CE was made of a submicron composite consisting of 60 wt.% $(La_{0.8}Sr_{0.2})_{0.95}MnO_3$ (LSM) and 40 wt.% 10Sc1CeSZ, as reported elsewhere [14,18]. Characterization of the electrode materials and porous layers included X-ray diffraction (XRD), scanning and transmission electron microscopy (SEM/TEM) coupled with energy dispersive spectroscopy (EDS), and thermogravimetric analysis (TGA); description of the equipment and experimental procedures can be found in Refs. [14,18-20]. Typical anode microstructures are illustrated in Fig.2 (b-d).

### 3. Results and Discussion

Fig.3(a) presents one representative example of the room-temperature Raman spectrum collected from the interface between oxidized cermet anode and 10Sc1YSZ s electrolyte. The peak at approximately 515 cm$^{-1}$ corresponds to 1P (1 phonon) oscillations in NiO [21]. The bands at ~740, 860 and 1100 cm$^{-1}$ are associated with 2P oscillations in nickel oxide, whilst the ~1460 cm$^{-1}$ peak originates from magnon (2M) oscillations. The strongest peak at ~615 cm$^{-1}$ corresponds to the sum of two contributions: NiO and cubic zirconia. Notice that a similar situation is characteristic for another strong peak, ~1100 cm$^{-1}$. At the same time, the 2M band totally disappears at the Neel temperature (252 $^{\circ}$C) due to the transformation of antiferromagnetic nickel oxide into the cubic β-NiO polymorph; the intensity of 515, 740 and 860 cm$^{-1}$ peaks is insufficient for quantitative analysis of the redox kinetics at elevated temperatures when the noise level becomes high. Hence, the reduction degree of NiO at the interface can only be evaluated from the relative intensities of 615 and 1100 cm$^{-1}$ peaks after subtraction of the cubic $ZrO_2$ contribution. The latter can be done by subtracting the final spectrum, collected on total reduction of the cermet when all NiO is converted into metallic Ni. This transformation of the Raman spectra was made after each reduction cycle. Fig.3b shows the variations of NiO subspectra with time during reduction in flowing 50% $H_2$ - 50% $N_2$ gas mixture at 500 $^{\circ}$C.

Time dependences of the 615 cm$^{-1}$ signal intensity on reduction of the cermet anode at 400-600 $^{\circ}$C are presented in Fig.3(c). In the course of Raman measurements, each reduction cycle was followed by re-oxidation in air at 900 C. After subsequent cooling, reproducibility of the Raman spectra was monitored until complete stabilization at a given temperature. As expected, the cermet reduction rate tends to exponentially increase on heating. Another necessary comment is that reproducible kinetic parameters can only be obtained after initial redox cycling, accompanied with substantial morphological reconstruction of NiO (Fig.2, c and d). As an example, Fig.4 displays the Raman peak intensity variations with time for the first 3 cycles after anode preparation, at 500 $^{\circ}$C. The kinetics observed on initial reduction is much slower compared to the 2$^{nd}$ and 3$^{rd}$ cycles where the reaction rate becomes stable. This trend is clearly associated with the increased surface area, decreased particle size and increased porosity of NiO component; the particle size and distribution of stabilized $ZrO_2$ matrix in the cermet remains unchanged (Fig.2d).

The dependence of the NiO reduction time ($t_r$) on reciprocal temperature is presented in Fig.5(a). The apparent activation energy, 1.7±0.2 eV, is close to the activation energy of $^{63}$Ni isotope diffusion along grain boundaries in NiO [22]. This behavior well corresponds to the literature data [23-26] showing that the kinetics of nickel oxide transformation into metallic Ni at intermediate conversion degrees is essentially determined by the transport and reaction rates at the NiO | Ni boundaries. Such kinetics can be analyzed using the Avrami model [25,26]:

$$\ln(-\ln(1-\alpha)) = \ln k + n \ln t \qquad (1)$$

where $\alpha$ is dimensionless reaction degree, $k$ is the rate constant, and $n$ is the exponent reflecting microscopic mechanisms of the process. One example of the Avrami plot is shown in Fig.5(b). This curve exhibits three distinct regions. The initial, non-linear part is related to the transient process when hydrogen diffuses through porous anode; the non-linear behavior at the final stage can be ascribed to decreasing penetration depth of the laser beam excitation due to covering of the grain surface by metallic Ni. The linear intermediate region of the Avrami plot corresponds to the $n$ value of 1.33±0.01, again in excellent agreement with literature [23,26]. The relevant mechanism may be associated with the metal nuclei growth limited by Ni diffusion [25,26]. Therefore, under open-circuit conditions the kinetics of NiO reduction at the porous cermet anode | solid electrolyte interface is similar to that of the bulk reaction, as expected. These results validate the experimental approach proposed for the *in-situ* Raman spectroscopy studies.

**Acknowledgements**

This work was supported by the Ministry of Education and Science of the Russian Federation (Studies and Developments on Priorities of Science and Technology federal program, unique identifier of the agreement RFMEFI61014X0007)5

**FIGURE CAPTIONS**

Fig.1. Electrode configuration in standard planar fuel cells (a); and model cell for the studies of the solid electrolyte | anode interface based on optically transparent solid electrolyte membrane (b), and schematic drawings of the high-temperature holder of the model cells (c) and entire experimental setup for the Raman spectroscopy studies (d).

Fig.2. Laue pattern of the 10Sc1YSZ single crystal membrane (a), SEM micrograph of fractured model electrochemical cell (b), and SEM images of as-prepared NiO-10Sc1CeSZ composite anode prior to reduction (c) and after redox cycling (d).

Fig.3. Example of the room-temperature Raman spectrum collected from as-sintered anode | electrolyte interface before reduction (a), Raman spectra taken from the interface as a function of time during reduction at 500 °C (b), and time dependences of 610 cm$^{-1}$ peak intensity on reduction at 400-600 °C (c). For (b), the spectra are shown with subtracted final spectrum (see text).

Fig.4. Time dependences of the NiO peak intensity on anode reduction during the 1st, 2nd and 3rd redox cycles.

Fig.5. Temperature dependence of the total reduction time of NiO-10Sc1CeSZ composite anode (a) and characteristic shape of the Raman intensity vs. time dependence plotted in Avrami coordinates (b).

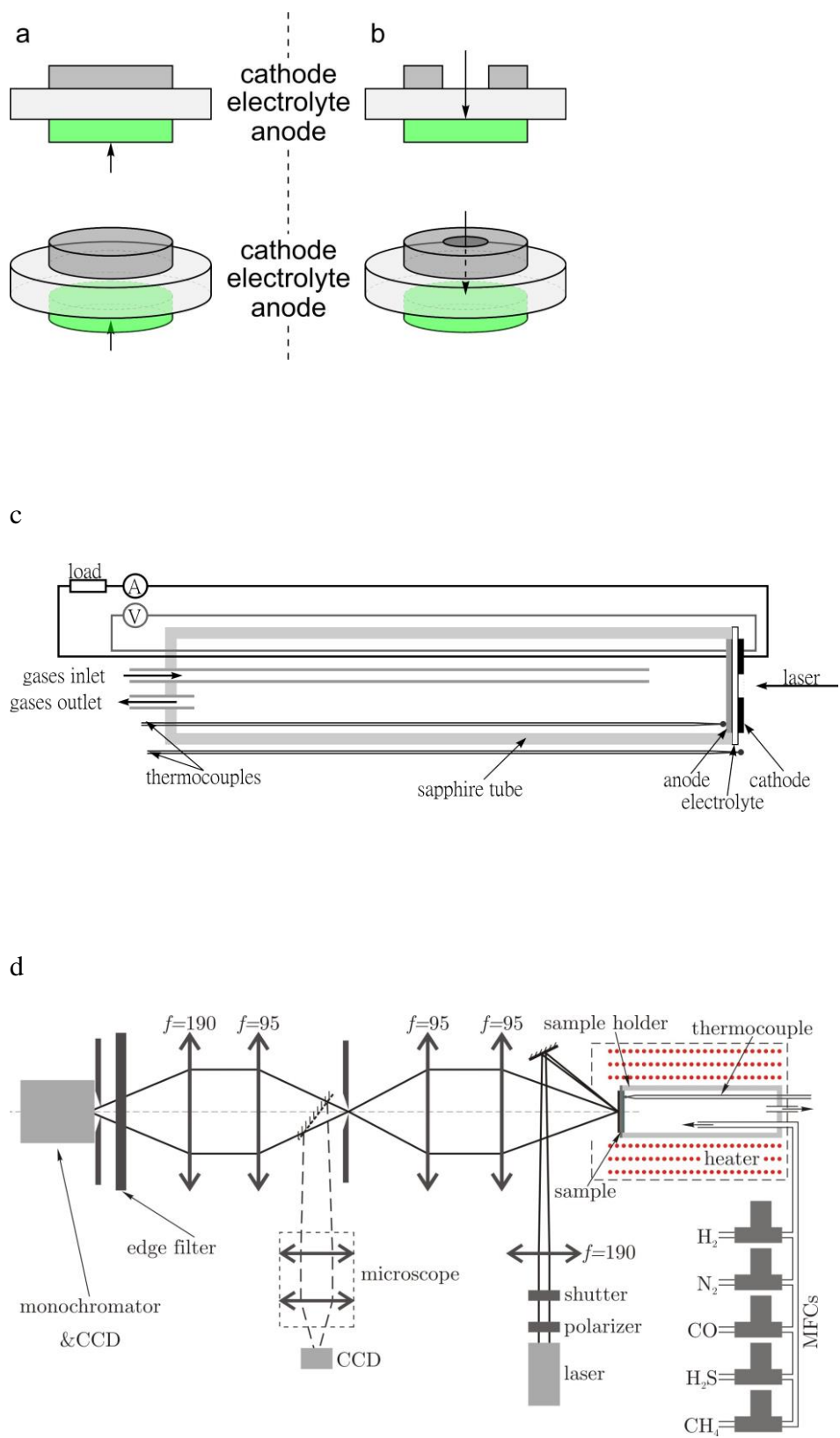

Fig.1

a 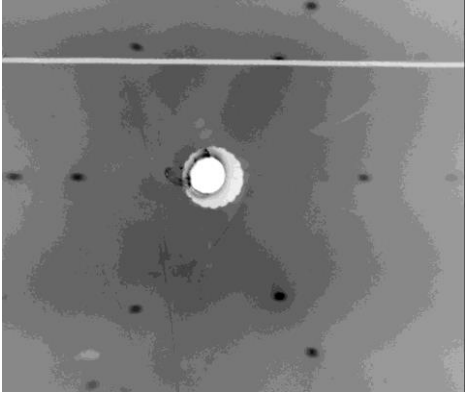 b 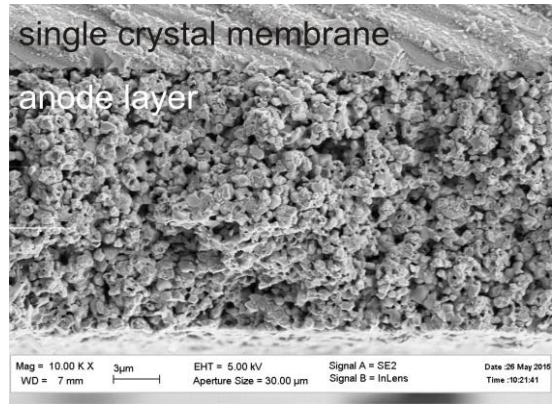

c 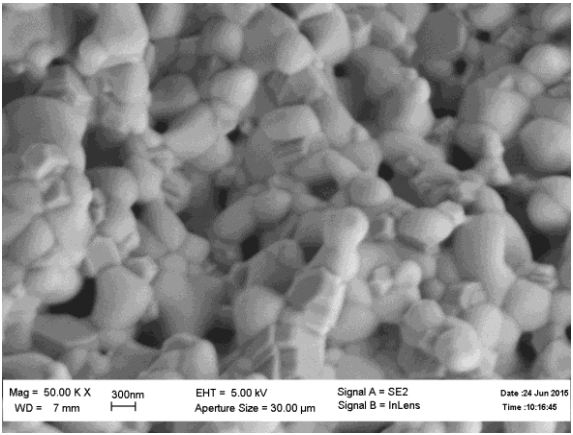 d 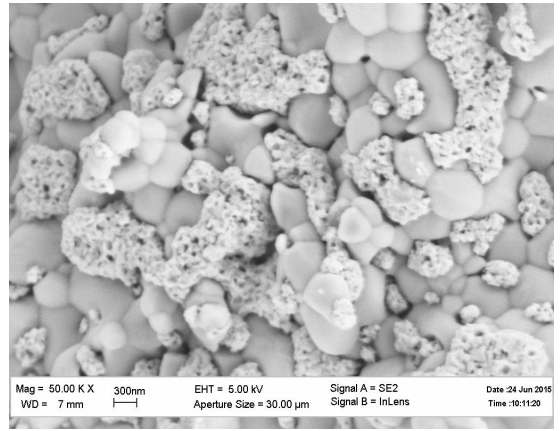

Fig.2

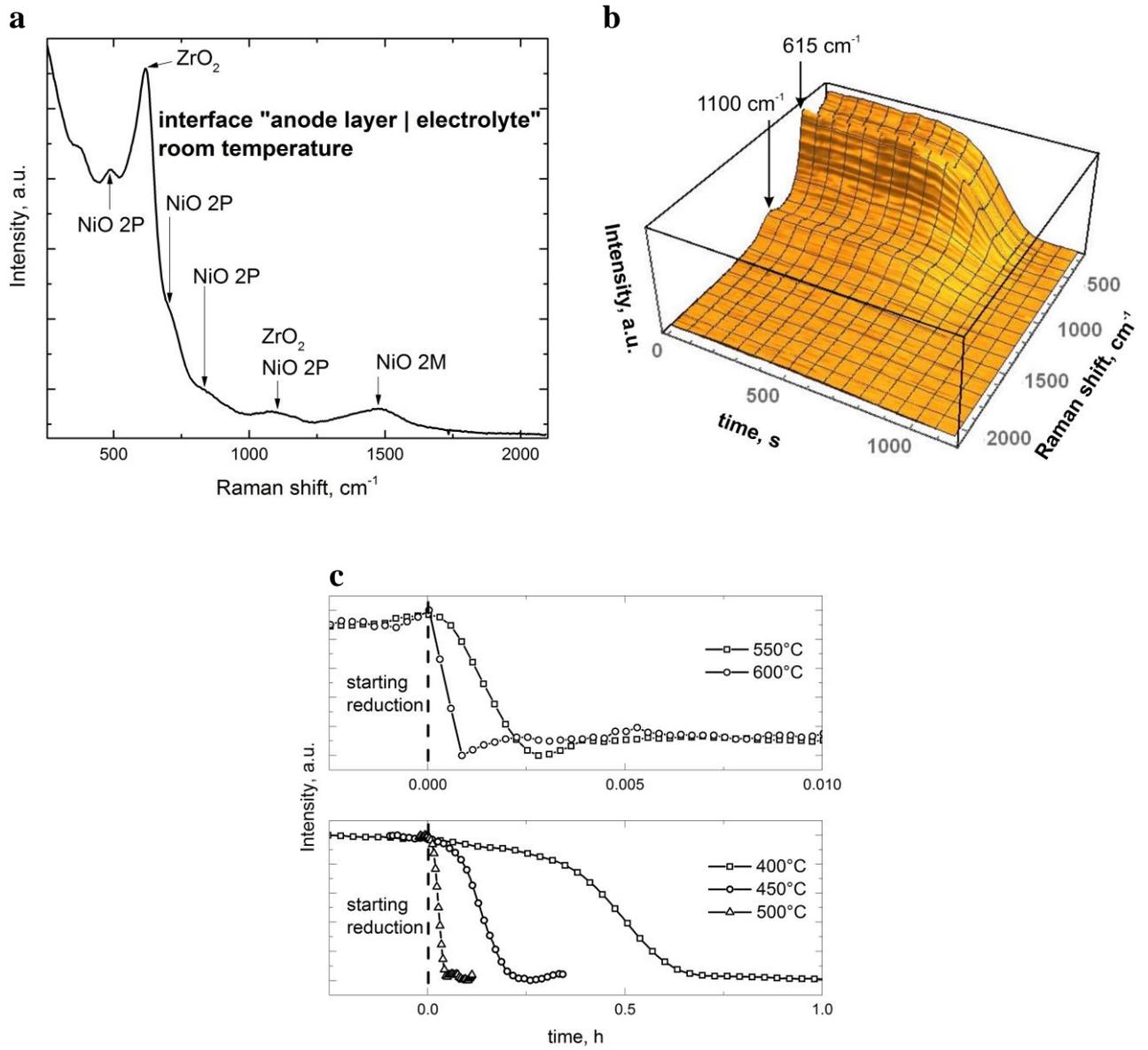

Fig.3

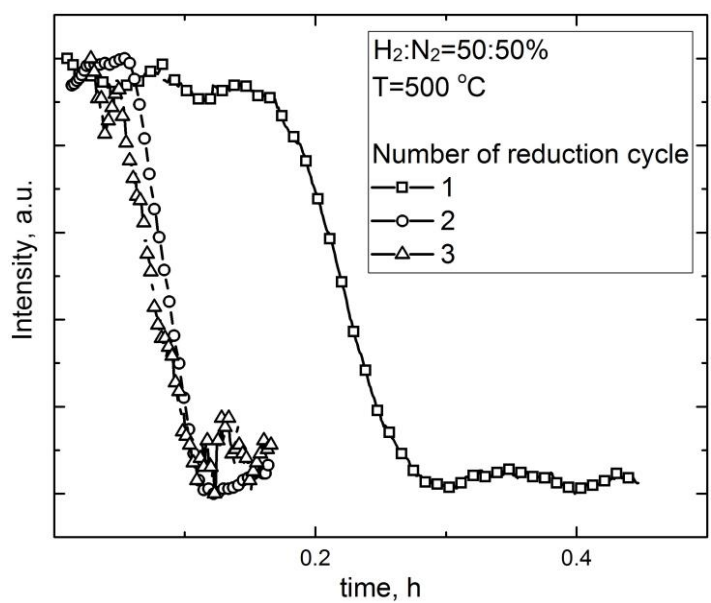

Fig.4

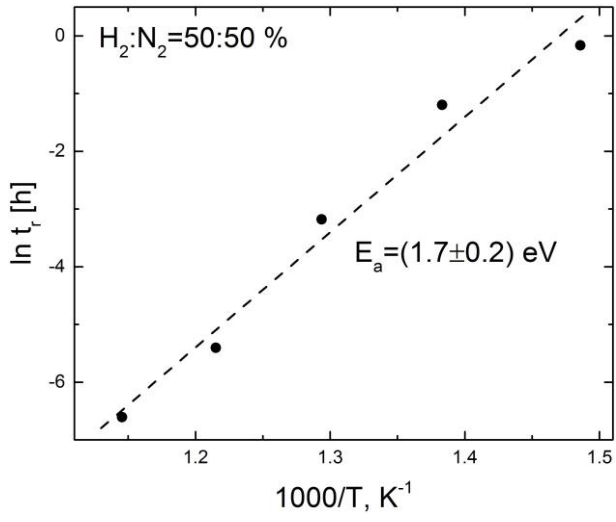

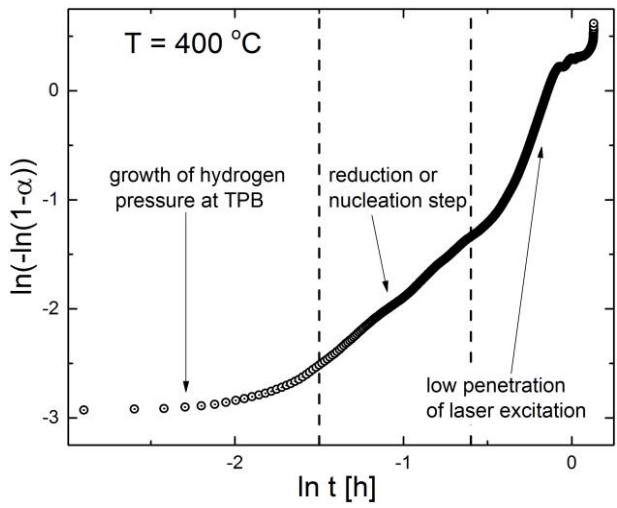

Fig.5